\documentclass[preprint,aps,amsmath,amssymb,showpacs,nofootinbib,superscriptaddress]{revtex4}
\usepackage{graphicx}
\usepackage[latin1]{inputenc}
\usepackage{dcolumn}
\usepackage{bm}

\newcommand{\br}{{\bf r}}

\begin{document}

\title{Graphene $n$-$p$ junction in a strong magnetic field: a
  semiclassical study}  
\author{Pierre Carmier} 
\affiliation{Univ.\ Paris-Sud, LPTMS UMR 8626, 91405 Orsay Cedex, France}
\author{Caio Lewenkopf}
\affiliation{Instituto de F\'{\i}sica, Universidade Federal
  Fluminense, 24210-346 Niter\'oi RJ, Brazil} 
\author{Denis Ullmo}
\affiliation{Univ.\ Paris-Sud, LPTMS UMR 8626, 91405 Orsay Cedex, France}
\affiliation{CNRS,  LPTMS UMR 8626, 91405   Orsay Cedex, France}    
\date{\today}

\begin{abstract}
We provide a semiclassical description of the electronic transport through graphene $n$-$p$ junctions in the quantum Hall regime. A semiclassical approximation for the conductance is derived in terms of the various snake-like trajectories at the interface of the junction. For a symmetric (ambipolar) configuration, the general result can be recovered by means of a simple scattering approach, providing a very transparent qualitative description of the problem under study. Consequences of our findings for the understanding of recent experiments are discussed.
\end{abstract}


\maketitle

Graphene is a two dimensional carbon based material which, due to its remarkable electronic and mechanical properties as well as potential applications, has been the subject of intense research activity in physics, chemistry and material sciences \cite{CastroNeto09,Geim09}. At the origin of this interest is the observation that the electronic low energy dynamics in graphene is governed by a Hamiltonian very similar to that of the 2-dimensional relativistic Dirac equation. This has a number of remarkable consequences, two of the most striking being an anomalous quantum Hall effect \cite{Novoselov05,Zhang05} and the existence of Klein tunneling \cite{Katsnelson06,Young09,Stander09}.

Experiments where both quantum Hall physics in graphene and Klein tunneling are at play were pioneered by Williams and collaborators, who measured the conductance of graphene $n$-$p$ junctions in a high magnetic field in the quantum Hall regime \cite{Williams07}. This was possible due to the deposition of a metallic topgate partially covering the graphene sheet. By independently varying the applied top and backgate voltages the resulting electrostatic potential creates negatively and positively doped regions in the graphene sample. In this way $n$-$n$, $p$-$p$, and $n$-$p$ junctions were produced. The two former did not show any striking feature. In contrast $n$-$p$ junctions, where the transition from the electron region to the hole region of the sample has to take place through Klein tunneling, showed a quite unexpected behavior.
Conductance plateaus at values $G_0/2$ and $3G_0/4$ ($G_0=2e^2/h$) were observed, at odds with the sequence $(2n+1)G_0$, $n \in \mathbb{Z}$, expected for the quantum Hall effect in graphene. Other experimental groups have observed these and other interesting plateaus in more elaborated set-ups \cite{Ozyilmaz07,Ki09,Lohmann09}. 

These observations were explained by Abanin and Levitov \cite{Abanin07sci} through a ``quantum chaos hypothesis'', which can be summarized as follows. At the interface between the $n$ and $p$ regions of the junction, the electrons experience a succession of Klein tunneling and skipping-orbit like propagation. It was  suggested that the mode mixing caused by this mechanism possibly leads the probability to be transmitted or reflected in a given mode to be perfectly ``democratic". The Landauer-B\"uttiker formula for the conductance under this hypothesis gives $G=G_0 N_n N_p/(N_n + N_p)$, where $N_n$ and $N_p$ are the number of channels in the $n$ and $p$ regions. This simple formula agrees with the observed conductance plateaus.

As already noted \cite{Abanin07sci}, this interpretation fails to provide a complete explanation of the experimental findings. Indeed, the ``quantum chaos hypothesis'' corresponds to the assumption that the scattering matrix describing mode mixing along the junction can be statistically modeled by random matrix theory. This hypothesis implies that the average scattering matrix coefficients are equal. However, it also predicts universal-conductance-like fluctuations \cite{Baranger94,Jalabert94}, which are quite robust and are not suppressed by adding disorder \cite{Long08,Li08}. Experiments show reasonably well defined plateaus, but no significant fluctuation amplitudes around a mean value. An alternative approach \cite{Tworzydlo07,Akhmerov08vv} showed that the edge boundary conditions affect the valley polarization of the zero energy Landau level. Clean edges produce new conductance plateaus in the ballistic regime, however not at the observed values.
The graphene $n-p$ junction plateaus remain an experimental mystery. The purpose of this Letter is to present a semiclassical analysis that provides a better understanding of mode mixing processes at the graphene $n$-$p$ interface regions, contributing to dispel the puzzle. 

\begin{figure}[]
\begin{center}
\includegraphics[width=1.0\linewidth]{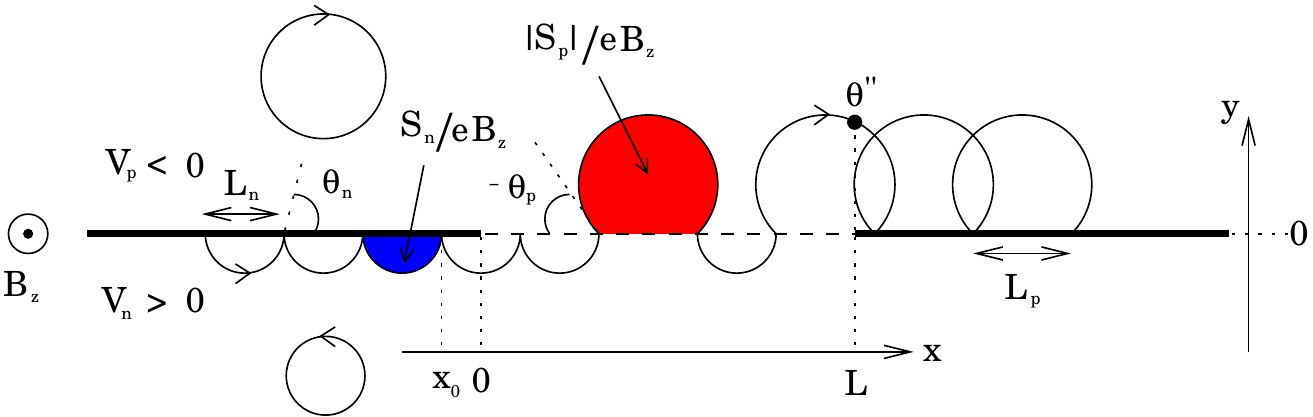}
\caption{(Color online) Scattering along a boundary intercalated with a step-like $n$-$p$ junction. The different symbols are explained in the text.}
\label{FigSnake}
\end{center}
\end{figure}

Mode mixing depends on several system-specific features: presence or absence of disorder, steepness of the potential barrier, possible diffractive effects when the electrons transit from their skipping orbit motion along the sample edges to the snake-like motion along the junction interface, etc. To specifically focus on the mixing occurring at the interface, we consider the model sketched in Fig.~\ref{FigSnake}: a perfectly clean graphene sample, the edges of which match smoothly with a step-like potential interface. The Landauer-B\"uttiker conductance $G$ across the junction is expressed as
\begin{equation}
\label{eq:cond}
G = G_0 \sum_{\kappa,\alpha} T_{\kappa,\alpha} \; ,
\end{equation}
where the sum runs over all the incoming skipping modes ${\kappa,\alpha}$ ($\alpha$ is the valley index) in for instance the $n$ region, and the $T_{\kappa,\alpha}$ are the corresponding transmission probabilities across the junction, for which we present a semiclassical description.

It is useful to begin by analyzing what are the results expected from a simple classical point of view. Let us call $V_n > 0$ and $V_p < 0$ the electrostatic potential in the $n$ and $p$ regions respectively (we assume the chemical potential $\mu =0$), $B_z$ the applied magnetic field, $\bf A$ the corresponding vector potential, and $v_F$ the Fermi velocity. In the interior of either regions, electrons experience a cyclotronic motion of opposite direction and of radius $R_{n,p} = |V_{n,p}|/(eB_z v_F)$. At the sample edge, the classical trajectories follow the skipping motion illustrated on Fig.~\ref{FigSnake} and can be characterized by the angle at which they hit the boundary.
We call this angle $\theta_n$ for trajectories in the $n$ region, and $\pi -\theta_p$ for trajectories in the $p$ region (as it turns out to be more convenient to label the direction of ${\bf \Pi} = {\bf p} + e {\bf A}({\bf r})$, which is antiparallel to the velocity in the hole region). Between two successive bounces, trajectories progress a distance $L_{n,p} = 2R_{n,p}\sin{\theta_{n,p}}$ along the edge.  The motion is similar at the $n$-$p$ interface, except that at each encounter with the potential barrier there is a probability
\begin{equation}
\label{eq:localT}
{\cal T} = \frac{ \sin{\theta_n} \sin{\theta_p} }{ \cos^2{ (
    \frac{\theta_n - \theta_p}{2} ) } } 
\end{equation} 
to be transmitted from the electron side to the hole side (or reciprocally) between two trajectories fulfilling the Snell-Descartes relation   
\begin{equation} 
\label{eq:snell} R_n \cos{\theta_n} = R_p \cos{\theta_p} \; .
\end{equation}

Let us consider an incoming mode (${\kappa,\alpha}$) from the $n$ region. Semiclassically, the mode is built on a one-parameter family of skipping trajectories corresponding to some fixed $\theta_n^{\kappa,\alpha}$. The different trajectories within the family are labeled by the abscissa $x_0 \in [-L_n(\theta_n^{\kappa,\alpha}),0]$ at which they last bounce before entering the interface region. The evolution of this set of trajectories under the above classical dynamics can be characterized by the probabilities $u_n(x)$ and $u_p(x)$ to emerge at $x$ on the electron side (with angle $\theta_n^{\kappa,\alpha}$) or on the hole side (with the corresponding angle given by Eq.~(\ref{eq:snell})). If the width $L$ of the junction is significantly larger than $L_n$ and
$L_p$, it can be shown that $u_n(x)$ and $u_p(x)$ converge both toward 1/2 exponentially quickly with the number of bounces on the junction. As a consequence, since whether a trajectory is transmitted or reflected only depends on the side it emerges from after the last scattering at the interface, the classical probability of transmission is, for a large junction, given by the ratio $L_p/(L_n+L_p)$. It should be stressed that this classical transmission probability, even if averaged on the angle $\theta_n^{\kappa,\alpha}$ specifying the incoming mode, does not correspond to what a ``quantum chaos hypothesis'' would require, namely a probability given by the proportion of available classical phase space on each side of the junction.

Let us now turn to the description of the quantum transmission. For the sake of definitiveness we discuss the case of infinite mass boundary at the edge of the graphene sample. We expect no qualitative differences for either armchair or zigzag boundary conditions. Let us introduce the actions 
\begin{eqnarray}
S_n(\theta_n) & = & e B_z R^2_n \left( \theta_n - \frac{ \sin{2\theta_n} }{2} \right) \; ,
\\
S_p(\theta_p) & = & - e B_z R^2_p \left( \pi - \theta_p + \frac{ \sin{2\theta_p} }{2} \right) \; .
\end{eqnarray}
Semiclassically, the edge modes in the $n$ region are built, within the leads, on the skipping trajectories bouncing on the boundary with an angle $\theta_{\kappa,\alpha}^{n}$ fulfilling the quantization condition
\begin{equation} 
\label{eq:quantize}
S_n(\theta_n^{\kappa,\alpha})= 2\pi\hbar (\kappa - \alpha /4)
\end{equation}
(again, $\alpha = \pm 1$ is the valley index). Note there is no $\kappa=0$ level for $\alpha=+1$.

\begin{figure}[]
\begin{center}
\includegraphics[width=0.3\linewidth]{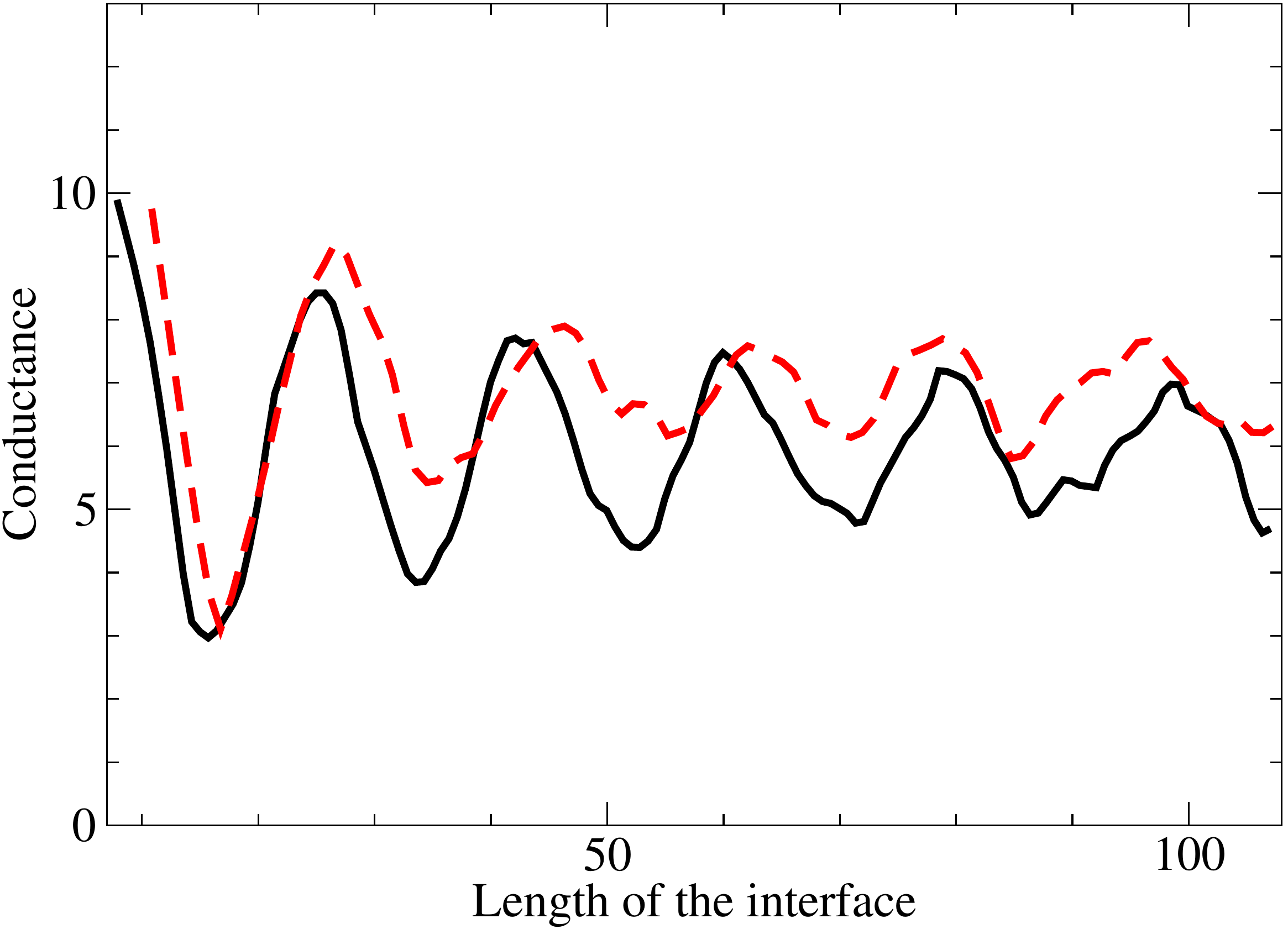}
\includegraphics[width=0.3\linewidth]{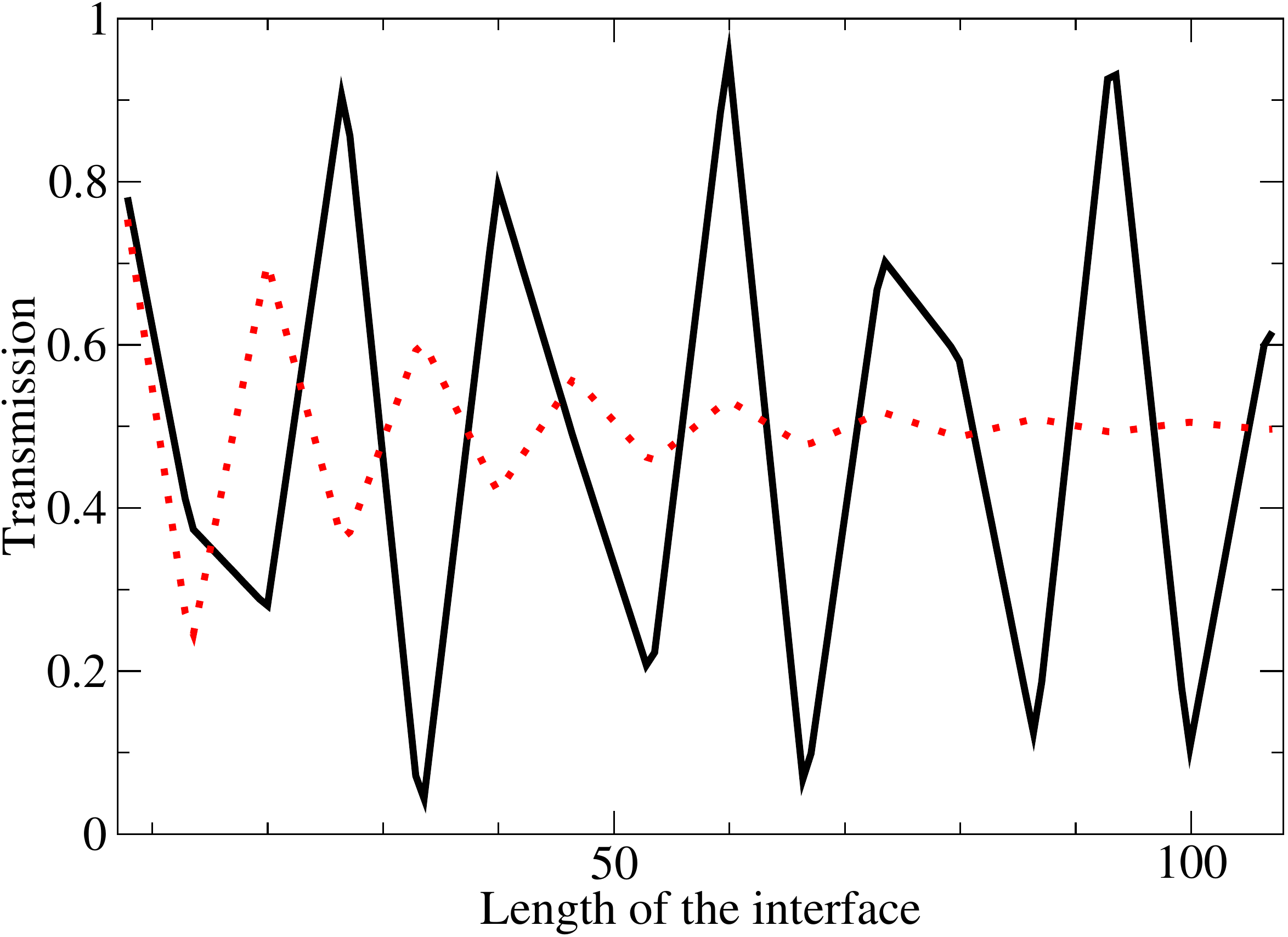}
\includegraphics[width=0.3\linewidth]{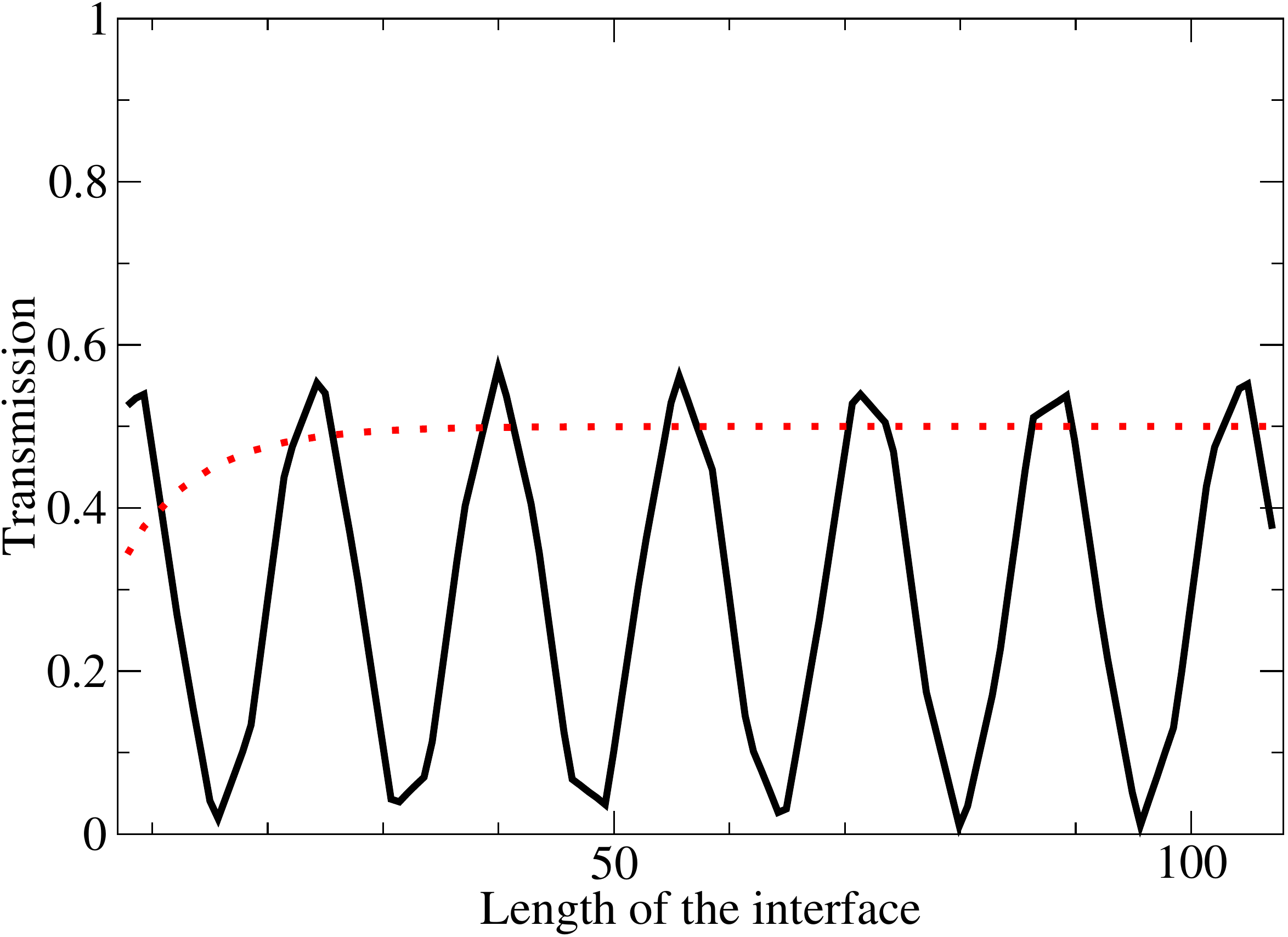}
\caption{(Color online) {\bf Left}: Conductance (in units of $G_0$) as a function of the length of the interface $L$ (in units of the magnetic length $l_B = \sqrt{\hbar/(eB_z)}$) for the case of 13 edge channels (filling factor 7) in the ambipolar (thick black line) and generic (red dashed line) cases. Evidently, no plateau in the conductance is reached in either cases. {\bf Middle and right}: Transmission of edge channels number 4 and 13, for which the local transmission probability in Eq.~(\ref{eq:localT}) is respectively ${\cal T}_4 = 0.87$ and ${\cal T}_{13} = 0.19$, for semiclassical (thick black line) and classical (red dotted line) dynamics. Substantial difference between both behaviors indicates strong influence of interferences between classical trajectories.}
\label{FigCond}
\end{center}
\end{figure}

We start our semiclassical discussion with the particularly simple case of ambipolar junctions, namely junctions such that $V_p = -V_n$. In that situation,  the Snell relation (\ref{eq:snell}) tells us that $\theta_n = \theta_p$ for any pair of angles, and therefore, $L_n(\theta_n) = L_p(\theta_p)$. As a consequence, the different trajectories within the family $\{ \theta_n = \theta^{\kappa,\alpha}_{n},x_0 \in [-L_n(\theta_n^{\kappa,\alpha}),0] \}$ follow a completely independent history. The propagation of the corresponding amplitudes along the junction can be performed for each of them separately and is obtained from the successive multiplications of two $2 \times 2$ unitary matrices. The first one,
\begin{equation}
P = \left( \begin{array}{cc} e^{\frac{i}{\hbar}S_n(\theta_n) - i\frac{\pi}{2} + i\xi} & 0 
\\ 
0 & e^{\frac{i}{\hbar}S_p(\theta_p) + i\frac{\pi}{2} + i\xi} \end{array} \right) \; ,
\end{equation}
with here $\theta_n = \theta_p = \theta^{\kappa,\alpha}_{n}$, describes the propagation on the electron and hole side between two interactions with the interface. The $\pm \pi/2$ are the Maslov phases originating from the focal point met on these pieces of trajectory, and $\xi = -\theta_n$ is a Berry phase. The second matrix describes the transmission and reflection between the electron and hole sides. It can be expressed as (in a form not restricted to the ambipolar case)
\begin{equation} 
\label{eq:S2}
D = \left( \begin{array}{cc} r_n e^{i \theta_n} & t_p e^{i (\theta_n+\theta_p)/2} 
\\  
t_n e^{i (\theta_p+\theta_n)/2}  & r_p e^{i \theta_p} \end{array} \right) \; ,
\end{equation} 
with $r_n = r_p = -{\cos  ((\theta_n + \theta_p)/2) }/{\cos((\theta_n - \theta_p)/2) }$ and $t_{n,p} = -i\sin{\theta_{n,p}} / \cos((\theta_n - \theta_p)/2) $. Note the phases $e^{i(\theta_{n,p} + \theta_{n,p})/2}$ can be interpreted also as Berry phases. The transmission probability for a given trajectory is then given by $T(\theta_n^{\kappa,\alpha},x_0) = \left| (0,1) (PD)^N (1,0)^T\right|^2$ with $N$ the integer part of $(L+|x_0| )/L_n$, which therefore depends on $x_0$. The total transmission for the mode $(\kappa,\alpha)$ is then obtained as $ T_{\kappa,\alpha} = L_n^{-1}\int_{-L_n}^0 dx_0 T(\theta_n^{\kappa,\alpha},x_0)$. The resulting conductance after summation on the modes $(\kappa,\alpha)$ is shown on Fig.~\ref{FigCond}.

In the generic case of arbitrary $V_n$ and $V_p$, a semiclassical expression for the conductance can be derived from the Fisher-Lee-Baranger-Stone equations \cite{Fisher81,Baranger89} relating the transmission and reflection coefficients $T_{\kappa,\alpha}$ and $R_{\kappa,\alpha}$ to the Green function
${\cal G}({\bf r''}, {\bf r'})$ and the incoming mode $\chi_{\kappa,\alpha}^{-}({\bf r'})$. In our case the transmission can be written as
\begin{equation}
T_{\kappa,\alpha} =  \int_{y''>0} dy'' t_{\kappa,\alpha}^{\dagger}({\bf
  r''})  v_F \sigma_x  t_{\kappa,\alpha}({\bf r''}) \; , 
\label{eq:Tka}
\end{equation}
with
\begin{equation}
t_{\kappa,\alpha} ({\bf r''}) = -i\hbar \int_{y'<0} dy' 
{\cal G}({\bf r''}, {\bf r'})  
v_F \sigma_x \chi_{\kappa,\alpha}^{-}({\bf r'}) \; . 
\label{eq:tka}
\end{equation}
The same results holds for $R_{\kappa,\alpha}$ except that the integral on $y''$ should be taken for negative values of $y''$ (i.e. in the electron region). The incoming mode is semiclassically expressed as
\begin{equation}
\label{modesemi}
\chi^{-}_{\kappa,\alpha} ({\bf r}) = \frac{ A_{\kappa,\alpha} e^{
    ik^{\kappa,\alpha}_x x} } {\sqrt{|\sin{\theta(y)}|}}
\sum_{\nu=\pm 1} e^{\frac{i}{\hbar}\nu S_{\kappa,\alpha}(y) - i\nu\frac{\pi}{4} }
 \left(
  \begin{array}{c} e^{-i\frac{\nu}{2}\theta(y)} \\  e^{i\frac{\nu}{2}\theta(y)}
  \end{array} \right) 
\end{equation}
with $\theta(y)$ the angle ($>0$) such that $y=R_n (\cos\theta_n^{\kappa,\alpha} - \cos\theta(y))$, $S_{\kappa,\alpha}(y) \equiv ({eB_0 R^2_n}/{2})(\theta(y) - \sin(2\theta(y))/2) $, and the normalization coefficient $ A_{\kappa,\alpha} = (4v_F R_n\sin{\theta_n^{\kappa,\alpha}})^{-1/2}$. For the Green function we take the semiclassical approximation derived in [\onlinecite{Carmier08}] for the graphene Hamiltonian, properly modified to account for
Klein tunneling at the junction interface. The semiclassical Green function is expressed as a sum over all trajectories $j$ joining $\br'$ to $\br''$ 
\begin{equation}
\label{semGrgra}
{\cal G}_{sc}({\bf r''},{\bf r'};E) = \frac{1}{i\hbar\sqrt{2\pi
    i\hbar}} \sum_j \left( \prod_{\zeta=n,p} \prod_{\gamma_\zeta,\eta_\zeta}
r_\zeta^{(\gamma_\zeta)}  t_\zeta^{(\eta_\zeta)} \right)
\frac{ e^{\frac{i}{\hbar} S_j({\bf r''},{\bf r'}) -
    i\frac{\pi}{2}\mu_j + i\xi_j}}{\sqrt{| J_j ({\bf r''},{\bf r'})
    |}}  
V^{\epsilon({\bf r''})}_j({\bf r''})V^{\epsilon({\bf r'})
  \dagger}_j({\bf r'}) \; , 
\end{equation}
where $S_j$ is the action integral along the orbit $j$, $J_j$ the stability determinant, $\mu_j$ the Maslov index counting the focal points met by the trajectory (counted negatively on the hole side), $\xi_j$ is a Berry phase equal to half of the angle of rotation of the vector ${\bf \Pi} = {\bf p} + e {\bf A}({\bf r})$ (which is parallel to the velocity on the electron side and antiparallel to it on the hole side), $V^+ = \frac{1}{\sqrt{2}} \binom{1}{e^{i \theta}}$, $V^- = \frac{1}{\sqrt{2}} \binom{e^{-i \theta}}{-1}$, with $\theta$ the direction of $\bf \Pi$, and $\epsilon({\bf r}) = +1$ or $-1$ depending
on whether $\bf r$ is on the electron or hole side. The only difference introduced by the presence of the potential barrier \cite{Couchman92} is that bounces on the interface, indexed by $\gamma_{n,p}$, as well as transmission across it, indexed by $\eta_{n,p}$, are taken into account through the coefficients $r_{n,p}$, $t_{n,p}$ introduced in Eq.~(\ref{eq:S2}) (in the definition of which the angles $\theta^{(\gamma_{n,p},\eta_{n,p})}$ are used).

A typical skipping orbit fulfilling the quantization condition Eq.~(\ref{eq:quantize}) is displayed on Fig.~\ref{FigSnake}. The orbit is characterized by the angle $\theta''$ at which it arrives at the Poincar\'e section $x=L$, the number of excursions $m_p$ on the hole side, and the number of traversals $k$ of the interface. The number of excursions on the electron side is then $m_n(m_p,\theta'') = \left[ (W - L'' - 2 m_p{R_p\sin{\theta_p}})/({2R_n\sin{\theta_n}}) \right]$ with $[\cdot]$ the integer part, and $L'' = R_p(\sin{\theta_p} - \sin{\theta''})$ for transmission and $R_n(\sin{\theta_n} + \sin{\theta''})$ for reflection. The trajectory should also be characterized by the ordering of the various excursions in the $n$ and $p$ sides of the junction. However these various orderings will correspond to the same amplitude, and therefore just contribute as a degeneracy factor given by $\Omega(m_p,k \! = \! 2k') = \binom{m_p -1}{k'-1} \binom{m_n(m_p) +1}{k'}$ for reflection, and $\Omega(m_p,k \! = \! 2k''+1) = \binom{m_p}{k''} \binom{m_n(m_p)}{k''}$ for transmission. Performing the integral in Eq.~(\ref{eq:tka}) in the stationary phase approximation, we get 
\begin{eqnarray}
T_{\kappa,\alpha} & = & \frac{R_p}{2 R_n\sin{\theta_n^{\kappa,\alpha}}} \int_{-\theta_T}^{\theta_T} d \theta'' \cos{\theta''} {\cal T}_{\kappa,\alpha}(\theta'') \; ,
\label{tmnfinal} 
\\
R_{\kappa,\alpha} & = & \frac{1}{2\sin{\theta_n^{\kappa,\alpha}}} \int_{-\theta_R}^{\theta_R} d \theta'' \cos{\theta''} {\cal R}_{\kappa,\alpha}(\theta'')
\label{rmnfinal} 
\end{eqnarray}
with

\begin{eqnarray}
{\cal T}_{\kappa,\alpha}(\theta'') & = & \left| t_n \sum_{m_p=0}^{M_p(\theta'')} (-i r_n e^{\frac{i}{\hbar}S_n(\theta_n^{\kappa,\alpha})})^{m_n(m_p,\theta'')} (ir_p e^{\frac{i}{\hbar}S_p(\theta_p)})^{m_p} \sum_{k''} \left(\frac{t_n t_p}{r_n r_p}\right)^{k''} \Omega(m_p,2k''+1) \right|^2 \; ,
\label{tmnfinal2} 
\\
{\cal R}_{\kappa,\alpha}(\theta'') & = & \left| \sum_{m_p=0}^{M_p(\theta'')} (-i r_n e^{\frac{i}{\hbar}S_n(\theta_n^{\kappa,\alpha})})^{m_n(m_p,\theta'')+1} (i r_p e^{\frac{i}{\hbar}S_p(\theta_p)})^{m_p} \sum_{k'} \left(\frac{t_n t_p}{r_n r_p}\right)^{k'} \Omega(m_p,2k') \right|^2
\label{rmnfinal2}
\end{eqnarray}
where $M_p(\theta'') = \left[ (W - L'')/({2R_p\sin{\theta_p}}) \right] $ and $\theta_T = \min(\theta_p,\pi-\theta_p)$, $\theta_R = \min(\theta_n^{\kappa,\alpha},\pi-\theta_n^{\kappa,\alpha})$. The corresponding curves for the conductance in Eq.~(\ref{eq:cond}) are shown on Fig.~\ref{FigCond}.

A few comments are in order. First, we stress that in the ambipolar case $V_p = -V_n$, the results obtained within the Baranger-Stone framework are, as expected, strictly identical to those derived earlier with the ``scattering matrix'' approach. Second, the equilibration and thus the saturation of the conductance expected at the classical level has to be contrasted with the persisting large
oscillations of the semiclassical conductance. Finally and more surprisingly, we observe that the mean of the semiclassical prediction can differ significantly from the classical limiting value. This is illustrated, for instance, at the rightmost panel of Fig.~\ref{FigCond}. 

Understanding the oscillating pattern in the conductance as a function of the length of the interface in Fig.~\ref{FigCond} is relatively straightforward in the ambipolar case, since the ``scattering matrix'' can be interpreted as that of a rotation on the Bloch sphere, acting on the vector composed of the reflection and transmission probability amplitudes of a given mode. In particular convergence to a given transmission probability, corresponding to a vector pointing at a fixed latitude, cannot be achieved. Outside of the ambipolar case, this interpretation is no longer valid but the pattern of the conductance in Fig.~\ref{FigCond} is similar enough to let us believe this picture remains qualitatively correct.

The advantage of a semiclassical description, compared for instance to
an exact numerical calculation of the conductance using recursive
Green function techniques, is that it makes it possible to discuss the
expected consequence of various modifications of the model we
consider. For instance we do not expect that including a finite width
$d_w$ in the potential step (as long as $d_w \ll l_B$) or changing the
edge boundary conditions will qualitatively modify the oscillating
pattern of the conductance. In the first case, the local probability
transmission Eq.~(\ref{eq:localT}) will be somewhat smaller, actually
improving the speed of classical convergence. In the second case, the
quantization condition in Eq.~(\ref{eq:quantize}) will be modified
\cite{Rakyta09}, but without affecting the basic mechanism at play
here. A geometry closer to the one used in experiments, i.e. with the
junction perpendicular to the edge of the ribbon, would imply some
diffraction at the edge-junction corner (this aspect will be discussed
in more details in [\onlinecite{Carmier10b}]). This, as more generally
the inclusion of a weak disorder, would somewhat diminish the
amplitude of the conductance oscillations and bring the average of the
semiclassical results closer to the classical conductance. 

As increasing the amount of disorder could only bring the system
toward a chaotic limit, characterized by universal conductance
fluctuations around a mean given by the classical (democratic)
expectation, our study indicates that within a model of perfectly
coherent electrons, the intrinsic properties of the $n$-$p$ junction
cannot produce the experimentally observed plateaus. These
considerations suggest the existence of inelastic processes occurring
in the vicinity of the junction, possibly reducing the coherence
length $\ell_\phi$ to values smaller than $L$. Further experiments,
varying the ratio $\ell_\phi/L$, are expected to provide further
insight on this issue.

We acknowledge a fruitful discussion with Alfredo Ozorio de
Almeida. This research was supported by the CAPES/COFECUB (project Ph
606/08).


\begin{thebibliography}{24}
\expandafter\ifx\csname natexlab\endcsname\relax\def\natexlab#1{#1}\fi
\expandafter\ifx\csname bibnamefont\endcsname\relax
  \def\bibnamefont#1{#1}\fi
\expandafter\ifx\csname bibfnamefont\endcsname\relax
  \def\bibfnamefont#1{#1}\fi
\expandafter\ifx\csname citenamefont\endcsname\relax
  \def\citenamefont#1{#1}\fi
\expandafter\ifx\csname url\endcsname\relax
  \def\url#1{\texttt{#1}}\fi
\expandafter\ifx\csname urlprefix\endcsname\relax\def\urlprefix{URL }\fi
\providecommand{\bibinfo}[2]{#2}
\providecommand{\eprint}[2][]{\url{#2}}

\bibitem[{\citenamefont{Neto et~al.}(2009)\citenamefont{Neto, Guinea, Peres,
  Novoselov, and Geim}}]{CastroNeto09}
\bibinfo{author}{\bibfnamefont{A.~H.~C.} \bibnamefont{Neto}},
  \bibinfo{author}{\bibfnamefont{F.}~\bibnamefont{Guinea}},
  \bibinfo{author}{\bibfnamefont{N.~M.~R.} \bibnamefont{Peres}},
  \bibinfo{author}{\bibfnamefont{K.~S.} \bibnamefont{Novoselov}},
  \bibnamefont{and} \bibinfo{author}{\bibfnamefont{A.~K.} \bibnamefont{Geim}},
  \bibinfo{journal}{Reviews of Modern Physics} \textbf{\bibinfo{volume}{81}},
  \bibinfo{pages}{109} (\bibinfo{year}{2009}).

\bibitem[{\citenamefont{Geim}(2009)}]{Geim09}
\bibinfo{author}{\bibfnamefont{A.~K.} \bibnamefont{Geim}},
  \bibinfo{journal}{Science} \textbf{\bibinfo{volume}{324}},
  \bibinfo{pages}{1530} (\bibinfo{year}{2009}).

\bibitem[{\citenamefont{Novoselov et~al.}(2005)\citenamefont{Novoselov, Geim,
  Morozov, Jiang, Katsnelson, Grigorieva, Dubonos, and Firsov}}]{Novoselov05}
\bibinfo{author}{\bibfnamefont{K.~S.} \bibnamefont{Novoselov}},
  \bibinfo{author}{\bibfnamefont{A.~K.} \bibnamefont{Geim}},
  \bibinfo{author}{\bibfnamefont{S.~V.} \bibnamefont{Morozov}},
  \bibinfo{author}{\bibfnamefont{D.}~\bibnamefont{Jiang}},
  \bibinfo{author}{\bibfnamefont{M.~I.} \bibnamefont{Katsnelson}},
  \bibinfo{author}{\bibfnamefont{I.~V.} \bibnamefont{Grigorieva}},
  \bibinfo{author}{\bibfnamefont{S.~V.} \bibnamefont{Dubonos}},
  \bibnamefont{and} \bibinfo{author}{\bibfnamefont{A.~A.}
  \bibnamefont{Firsov}}, \bibinfo{journal}{Nature}
  \textbf{\bibinfo{volume}{438}}, \bibinfo{pages}{197} (\bibinfo{year}{2005}).

\bibitem[{\citenamefont{Zhang et~al.}(2005)\citenamefont{Zhang, Tan, Stormer,
  and Kim}}]{Zhang05}
\bibinfo{author}{\bibfnamefont{Y.}~\bibnamefont{Zhang}},
  \bibinfo{author}{\bibfnamefont{Y.~W.} \bibnamefont{Tan}},
  \bibinfo{author}{\bibfnamefont{H.~L.} \bibnamefont{Stormer}},
  \bibnamefont{and} \bibinfo{author}{\bibfnamefont{P.}~\bibnamefont{Kim}},
  \bibinfo{journal}{Nature} \textbf{\bibinfo{volume}{438}},
  \bibinfo{pages}{201} (\bibinfo{year}{2005}).

\bibitem[{\citenamefont{Katsnelson et~al.}(2006)\citenamefont{Katsnelson,
  Novoselov, and Geim}}]{Katsnelson06}
\bibinfo{author}{\bibfnamefont{M.~I.} \bibnamefont{Katsnelson}},
  \bibinfo{author}{\bibfnamefont{K.~S.} \bibnamefont{Novoselov}},
  \bibnamefont{and} \bibinfo{author}{\bibfnamefont{A.~K.} \bibnamefont{Geim}},
  \bibinfo{journal}{Nature Physics} \textbf{\bibinfo{volume}{2}},
  \bibinfo{pages}{620} (\bibinfo{year}{2006}).

\bibitem[{\citenamefont{Young and Kim}(2009)}]{Young09}
\bibinfo{author}{\bibfnamefont{A.~F.} \bibnamefont{Young}} \bibnamefont{and}
  \bibinfo{author}{\bibfnamefont{P.}~\bibnamefont{Kim}},
  \bibinfo{journal}{Nature Physics} \textbf{\bibinfo{volume}{5}},
  \bibinfo{pages}{222} (\bibinfo{year}{2009}).

\bibitem[{\citenamefont{Stander et~al.}(2009)\citenamefont{Stander, Huard, and
  Goldhaber-Gordon}}]{Stander09}
\bibinfo{author}{\bibfnamefont{N.}~\bibnamefont{Stander}},
  \bibinfo{author}{\bibfnamefont{B.}~\bibnamefont{Huard}}, \bibnamefont{and}
  \bibinfo{author}{\bibfnamefont{D.}~\bibnamefont{Goldhaber-Gordon}},
  \bibinfo{journal}{Phys.~Rev.~Lett.} \textbf{\bibinfo{volume}{102}},
  \bibinfo{pages}{026807} (\bibinfo{year}{2009}).

\bibitem[{\citenamefont{Williams et~al.}(2007)\citenamefont{Williams, DiCarlo,
  and Marcus}}]{Williams07}
\bibinfo{author}{\bibfnamefont{J.~R.} \bibnamefont{Williams}},
  \bibinfo{author}{\bibfnamefont{L.}~\bibnamefont{DiCarlo}}, \bibnamefont{and}
  \bibinfo{author}{\bibfnamefont{C.~M.} \bibnamefont{Marcus}},
  \bibinfo{journal}{Science} \textbf{\bibinfo{volume}{317}},
  \bibinfo{pages}{638} (\bibinfo{year}{2007}).

\bibitem[{\citenamefont{Ozyilmaz et~al.}(2007)\citenamefont{Ozyilmaz,
  Jarillo-Herrero, Efetov, Abanin, Levitov, and Kim}}]{Ozyilmaz07}
\bibinfo{author}{\bibfnamefont{B.}~\bibnamefont{Ozyilmaz}},
  \bibinfo{author}{\bibfnamefont{P.}~\bibnamefont{Jarillo-Herrero}},
  \bibinfo{author}{\bibfnamefont{D.}~\bibnamefont{Efetov}},
  \bibinfo{author}{\bibfnamefont{D.~A.} \bibnamefont{Abanin}},
  \bibinfo{author}{\bibfnamefont{L.~S.} \bibnamefont{Levitov}},
  \bibnamefont{and} \bibinfo{author}{\bibfnamefont{P.}~\bibnamefont{Kim}},
  \bibinfo{journal}{Phys.~Rev.~Lett.} \textbf{\bibinfo{volume}{99}},
  \bibinfo{pages}{166804} (\bibinfo{year}{2007}).

\bibitem[{\citenamefont{Ki and Lee}(2009)}]{Ki09}
\bibinfo{author}{\bibfnamefont{D.-K.} \bibnamefont{Ki}} \bibnamefont{and}
  \bibinfo{author}{\bibfnamefont{H.-J.} \bibnamefont{Lee}},
  \bibinfo{journal}{Phys.~Rev.~B} \textbf{\bibinfo{volume}{79}},
  \bibinfo{pages}{195327} (\bibinfo{year}{2009}).

\bibitem[{\citenamefont{Lohmann et~al.}(2009)\citenamefont{Lohmann, von
  Klitzing, and Smet}}]{Lohmann09}
\bibinfo{author}{\bibfnamefont{T.}~\bibnamefont{Lohmann}},
  \bibinfo{author}{\bibfnamefont{K.}~\bibnamefont{von Klitzing}},
  \bibnamefont{and} \bibinfo{author}{\bibfnamefont{J.~H.} \bibnamefont{Smet}},
  \bibinfo{journal}{Nano Letters} \textbf{\bibinfo{volume}{9}},
  \bibinfo{pages}{1973} (\bibinfo{year}{2009}).

\bibitem[{\citenamefont{Abanin and Levitov}(2007)}]{Abanin07sci}
\bibinfo{author}{\bibfnamefont{D.~A.} \bibnamefont{Abanin}} \bibnamefont{and}
  \bibinfo{author}{\bibfnamefont{L.~S.} \bibnamefont{Levitov}},
  \bibinfo{journal}{Science} \textbf{\bibinfo{volume}{317}},
  \bibinfo{pages}{641} (\bibinfo{year}{2007}).

\bibitem[{\citenamefont{Baranger and Mello}(1994)}]{Baranger94}
\bibinfo{author}{\bibfnamefont{H.~U.} \bibnamefont{Baranger}} \bibnamefont{and}
  \bibinfo{author}{\bibfnamefont{P.~A.} \bibnamefont{Mello}},
  \bibinfo{journal}{Phys.~Rev.~Lett.} \textbf{\bibinfo{volume}{73}},
  \bibinfo{pages}{142} (\bibinfo{year}{1994}).

\bibitem[{\citenamefont{Jalabert et~al.}(1994)\citenamefont{Jalabert, Pichard,
  and Beenakker}}]{Jalabert94}
\bibinfo{author}{\bibfnamefont{R.~A.} \bibnamefont{Jalabert}},
  \bibinfo{author}{\bibfnamefont{J.-L.} \bibnamefont{Pichard}},
  \bibnamefont{and} \bibinfo{author}{\bibfnamefont{C.~W.~J.}
  \bibnamefont{Beenakker}}, \bibinfo{journal}{EPL (Europhysics Letters)}
  \textbf{\bibinfo{volume}{27}}, \bibinfo{pages}{255} (\bibinfo{year}{1994}).

\bibitem[{\citenamefont{Long et~al.}(2008)\citenamefont{Long, Sun, and
  Wang}}]{Long08}
\bibinfo{author}{\bibfnamefont{W.}~\bibnamefont{Long}},
  \bibinfo{author}{\bibfnamefont{Q.~F.} \bibnamefont{Sun}}, \bibnamefont{and}
  \bibinfo{author}{\bibfnamefont{J.}~\bibnamefont{Wang}},
  \bibinfo{journal}{Phys.~Rev.~Lett.} \textbf{\bibinfo{volume}{101}},
  \bibinfo{pages}{166806} (\bibinfo{year}{2008}).

\bibitem[{\citenamefont{Li and Shen}(2008)}]{Li08}
\bibinfo{author}{\bibfnamefont{J.}~\bibnamefont{Li}} \bibnamefont{and}
  \bibinfo{author}{\bibfnamefont{S.~Q.} \bibnamefont{Shen}},
  \bibinfo{journal}{Phys.~Rev.~B} \textbf{\bibinfo{volume}{78}},
  \bibinfo{pages}{205308} (\bibinfo{year}{2008}).

\bibitem[{\citenamefont{Tworzydlo et~al.}(2007)\citenamefont{Tworzydlo, Snyman,
  Akhmerov, and Beenakker}}]{Tworzydlo07}
\bibinfo{author}{\bibfnamefont{J.}~\bibnamefont{Tworzydlo}},
  \bibinfo{author}{\bibfnamefont{I.}~\bibnamefont{Snyman}},
  \bibinfo{author}{\bibfnamefont{A.~R.} \bibnamefont{Akhmerov}},
  \bibnamefont{and} \bibinfo{author}{\bibfnamefont{C.~W.~J.}
  \bibnamefont{Beenakker}}, \bibinfo{journal}{Phys.~Rev.~B}
  \textbf{\bibinfo{volume}{76}}, \bibinfo{pages}{035411}
  (\bibinfo{year}{2007}).

\bibitem[{\citenamefont{Akhmerov et~al.}(2008)\citenamefont{Akhmerov,
  Bardarson, Rycerz, and Beenakker}}]{Akhmerov08vv}
\bibinfo{author}{\bibfnamefont{A.~R.} \bibnamefont{Akhmerov}},
  \bibinfo{author}{\bibfnamefont{J.~H.} \bibnamefont{Bardarson}},
  \bibinfo{author}{\bibfnamefont{A.}~\bibnamefont{Rycerz}}, \bibnamefont{and}
  \bibinfo{author}{\bibfnamefont{C.~W.~J.} \bibnamefont{Beenakker}},
  \bibinfo{journal}{Phys.~Rev.~B} \textbf{\bibinfo{volume}{77}},
  \bibinfo{pages}{205416} (\bibinfo{year}{2008}).

\bibitem[{\citenamefont{Fisher and Lee}(1981)}]{Fisher81}
\bibinfo{author}{\bibfnamefont{D.~S.} \bibnamefont{Fisher}} \bibnamefont{and}
  \bibinfo{author}{\bibfnamefont{P.~A.} \bibnamefont{Lee}},
  \bibinfo{journal}{Phys.~Rev.~B} \textbf{\bibinfo{volume}{23}},
  \bibinfo{pages}{6851} (\bibinfo{year}{1981}).

\bibitem[{\citenamefont{Baranger and Stone}(1989)}]{Baranger89}
\bibinfo{author}{\bibfnamefont{H.~U.} \bibnamefont{Baranger}} \bibnamefont{and}
  \bibinfo{author}{\bibfnamefont{A.~D.} \bibnamefont{Stone}},
  \bibinfo{journal}{Phys.~Rev.~B} \textbf{\bibinfo{volume}{40}},
  \bibinfo{pages}{8169} (\bibinfo{year}{1989}).

\bibitem[{\citenamefont{Carmier and Ullmo}(2008)}]{Carmier08}
\bibinfo{author}{\bibfnamefont{P.}~\bibnamefont{Carmier}} \bibnamefont{and}
  \bibinfo{author}{\bibfnamefont{D.}~\bibnamefont{Ullmo}},
  \bibinfo{journal}{Phys.~Rev.~B} \textbf{\bibinfo{volume}{77}},
  \bibinfo{pages}{245413} (\bibinfo{year}{2008}).

\bibitem[{\citenamefont{Couchman et~al.}(1992)\citenamefont{Couchman, Ott, and
  Antonsen}}]{Couchman92}
\bibinfo{author}{\bibfnamefont{L.}~\bibnamefont{Couchman}},
  \bibinfo{author}{\bibfnamefont{E.}~\bibnamefont{Ott}}, \bibnamefont{and}
  \bibinfo{author}{\bibfnamefont{T.~M.} \bibnamefont{Antonsen}},
  \bibinfo{journal}{Phys.~Rev.~A} \textbf{\bibinfo{volume}{46}},
  \bibinfo{pages}{6193} (\bibinfo{year}{1992}).

\bibitem[{\citenamefont{Rakyta et~al.}(2009)\citenamefont{Rakyta, Kormanyos,
  and Cserti}}]{Rakyta09}
\bibinfo{author}{\bibfnamefont{P.}~\bibnamefont{Rakyta}},
  \bibinfo{author}{\bibfnamefont{A.}~\bibnamefont{Kormanyos}},
  \bibnamefont{and} \bibinfo{author}{\bibfnamefont{J.}~\bibnamefont{Cserti}},
  \emph{\bibinfo{title}{Exploring the graphene edges with coherent electron
  focusing}} (\bibinfo{year}{2009}),
  \urlprefix\url{http://www.citebase.org/abstract?id=oai:arXiv.org:0909.1705}.

\bibitem[{\citenamefont{Carmier et~al.}(2010)\citenamefont{Carmier, Lewenkopf,
  and Ullmo}}]{Carmier10b}
\bibinfo{author}{\bibfnamefont{P.}~\bibnamefont{Carmier}},
  \bibinfo{author}{\bibfnamefont{C.}~\bibnamefont{Lewenkopf}},
  \bibnamefont{and} \bibinfo{author}{\bibfnamefont{D.}~\bibnamefont{Ullmo}},
  \bibinfo{journal}{in preparation}  (\bibinfo{year}{2010}).

\end{thebibliography}

\end{document}